\documentclass[12pt,preprint]{article}
\usepackage{amssymb}
\usepackage{amsmath}
\usepackage{graphics}
\usepackage{epsfig}

\renewcommand{\vec}[1]{{\bf #1}}
\newcommand{\eqb}{\begin{equation}}
\newcommand{\eqe}{\end{equation}}
\newcommand{\dmb}{\begin{displaymath}}
\newcommand{\dme}{\end{displaymath}}
\newcommand{\pd}{\partial}

\newcommand{\eab}{\begin{eqnarray}}
\newcommand{\eae}{\end{eqnarray}}

\newcommand{\be}{\begin{equation}}
\newcommand{\ee}{\end{equation}}

\begin{document}

\begin{titlepage}
\begin{flushright} 
KA-TP-11-2007
\end{flushright}
\vspace{0.6cm}

\begin{center}
\Large{Irreducible three-loop contributions to the pressure 
in Yang-Mills thermodynamics}
\vspace{1.5cm}

\large{Dariush Kaviani$^\dagger$ and Ralf Hofmann$^*$}

\end{center}
\vspace{1.5cm} 
\begin{center}
{\em $\mbox{}^\dagger$ Institut f\"ur Theoretische Physik\\ 
Universit\"at Heidelberg\\ 
Philosophenweg 16\\ 
69120 Heidelberg, Germany}
\end{center}
\vspace{1.5cm}

\begin{center}
{\em $\mbox{}^*$ Institut f\"ur Theoretische Physik\\ 
Universit\"at Karlsruhe (TH)\\ 
Kaiserstr. 12\\ 
76131 Karlsruhe, Germany}
\end{center}
\vspace{1.5cm}

\begin{abstract}

In the effective theory for the deconfining phase of SU(2) Yang-Mills thermodynamics 
we compute estimates for the moduli of the irreducible three-loop diagrams 
contributing to the pressure. Our numerical 
results are in agreement with general expectations.

\end{abstract} 

\end{titlepage}

\section{Introduction}

To obtain essential analytical insights into the 
thermodynamics of four-dimensional Yang-Mills 
theories is difficult even at high temperature 
where gauge fields do propagate. A perturbative treatment of Yang-Mills thermodynamics, 
which is technically 
highly involved, runs into problems that are 
associated with the masslessness of the fundamental, propagating 
degrees of freedom. A manifestation of this problem is the 
apparent nonconvergence of the perturbative series\footnote{At $T=0$ it is suggestive that the perturbative
series represents an asymptotic expansion, and 
so the first few orders do capture a great deal of the physics \cite{tHooftVeltman,NP2004}. The perturbative
series at finite temperature apparently does not enjoy this property.} 
which can be pinned down to the weak screening in the 
magnetic sector of the theory \cite{Linde1980}. Loosely speaking, 
this problem arises because the perturbative a priori 
estimate for the thermal ground state is inappropriate. Namely, the presence of 
topologically nontrivial fluctuations, 
which do contribute to the thermodynamics of the Yang-Mills 
system\footnote{For example, 
the trace of the energy-momentum tensor rises linearly with temperature 
\cite{GH2007}, and the dimensionally reduced theory confines \cite{Korthals-Altes}.} 
in a direct (ground state) and indirect (masses) way, is neglected in perturbative 
loop expansions.        

In \cite{Hofmann} a nonperturbative approach to SU(2) and SU(3) Yang-Mills 
thermodynamics is developed. Let us discuss the deconfining phase for the 
SU(2) case only. The idea is to first derive a 
thermal ground state which is composed of interacting topological fluctuations: 
calorons and anticalorons. While the dynamics of these fluctuations would appear to be 
highly complex in the hypothetic case, where an externally provided probe of a given 
momentum transfer is applied to the system, the case of 
pure, infinite-volume thermodynamics\footnote{Subject to two scales only: temperature $T$ and Yang-Mills scale $\Lambda$.} 
selfconsistently adjusts a maximal resolution $|\phi|$ in 
such a way that the ground state admits a remarkably simple 
analytical description. To derive this situation 
a spatial coarse-graining needs to be performed which yet turns out to be sufficiently 
local to only require consideration of calorons and anticalorons of 
topological charge modulus $|Q|=1$ \cite{HerbstHofmann2004}. As a consequence, 
an adjoint and inert (spatially homogeneous) scalar field $\phi$ emerges which together 
with a pure-gauge (coarse-grained) configuration describes the 
ground state of the system, see also \cite{garfield}. 
The presence of $\phi$ signals a dynamical gauge-symmetry 
breaking SU(2)$\to$U(1) implying that two out of the three propagating and  
coarse-grained gauge-mode species acquire a (temperature-dependent) mass. 
This and the fact that the off-shellness of these modes and the momentum 
transfer in local vertices is highly constrained after 
spatial coarse-graining imply the rapid convergence of loop expansions 
in the effective theory \cite{Hofmann2006}. The purpose of the present 
article is to demonstrate this by estimating the irreducible three-loop 
contribution to the pressure in deconfining SU(2) Yang-Mills thermodynamics. 

The paper is organized as follows. In Sec.\,\ref{ETD} we briefly review technical 
essentials of the effective theory. Sec.\,\ref{T3D} first explains 
what is meant by the term `irreducible diagram' 
according to the discussion in \cite{Hofmann2006}. Subsequently, 
we elucidate the structure of irreducible three-loop diagrams. 
In Sec.\,\ref{res} we perform the integrations numerically using the 
Monte-Carlo method. Results are indicated and discussed. Finally, Sec.\,\ref{con} 
gives our conclusions.  

\section{Effective theory for deconfining phase\label{ETD}}

Here we very briefly review the effective theory for SU(2) Yang-Mills thermodynamics being in its deconfining
phase. The following effective action emerges upon a selfconsistent spatial 
coarse-graining involving interacting nontrivial-holonomy calorons of topological 
charge modulus $|Q|=1$ (ground state) and topologically trivial gauge fields (excitations) 
\cite{Hofmann}:
\eqb
\label{act}
S = \mbox{tr}\, \int_0^\beta d\tau \int d^3x \left( \frac12\,G_{\mu\nu}G_{\mu\nu} +
D_\mu \phi D_\mu \phi + \Lambda^6 \phi^{-2} \right)\,.
\eqe
In Eq.\,(\ref{act}) $G_{\mu\nu}\equiv G^a_{\mu\nu} \frac{\lambda^a}{2}$, 
$G^a_{\mu\nu}=\pd_\mu A^a_\nu-\pd_\nu A^a_\mu+e\,\epsilon^{abc}A^b_\mu A^c_\nu$, and 
$D_\mu \phi=\partial_\mu \phi + ie[\phi,A_\mu]$ 
where $A_\mu$ is the (coarse-grained) gauge field of 
trivial topology, and $e$ denotes the effective gauge coupling. 

In a first step, only a noninteracting (trivial-holonomy) caloron and anticaloron 
is coarse-grained over an infinite spatial volume into 
the phase of a spatially homogeneous adjoint scalar field 
$\phi$. Subsequently, the presence of a Yang-Mills scale $\Lambda$ is assumed\footnote{This assumption is
actually redundant since $\Lambda$ can be interpreted as a nonperturbative integration constant, see
\cite{garfield}.} to obtain the modulus 
$|\phi|=\sqrt{\frac{\Lambda^3}{2\pi T}}$. The latter determines the finite spatial 
length scale $|\phi|^{-1}$ at which the above coarse-graining saturates. 
Taking interactions between calorons and 
anticalorons into account, a (coarse-grained) pure-gauge configuration emerges. 
Going from a gauge, where $\phi$ 
winds along the compactified euclidean time dimension, to unitary gauge\footnote{
This involves an 
admissible electric center transformation \cite{Hofmann,SHG2006}.} we arrive at
\begin{equation}
\mathcal{L}_{{\tiny \mbox{eff}}}^{u.g.}=\mathcal{L}\left[ a_{\mu }\right] =%
\frac{1}{4}\left( G_{E}^{a,\mu \nu }[a_{\mu }]\right) ^{2}+2e^{2}\left\vert
\phi \right\vert ^{2}\left( \left( a_{\mu }^{1}\right) ^{2}+\left( a_{\mu
}^{2}\right) ^{2}\right) +2\frac{\Lambda ^{6}}{\left\vert \phi \right\vert
^{2}}\,.  \label{lagfirstord}
\end{equation}
The effective gauge coupling $e$ enters into both the effective field strength 
$G_{E}^{a,\mu \nu}$ and the mass $m$ 
for the fields $a_{\mu}^{1,2}$. One has 
\begin{equation}
m^{2}=m(T)^{2}=m_{1}^{2}=m_{2}^{2}=4e^{2}\left\vert \phi \right\vert
^{2}\,,\ \ \ \ m_{3}^{2}=0\,.  \label{masssp}
\end{equation}
The gauge mode $a_{\mu }^{3}$ stays massless on tree-level in the effective theory 
(adjoint Higgs mechanism). The associated unbroken U(1) gauge freedom is 
fixed by imposing the Coulomb condition $\pd_i a^3_i=0$. 
This corresponds to a completely fixed and {\sl physical} gauge (unitary-Coulomb gauge).   

Let us now give the one-loop expressions for the partial 
pressures $P_1=P_2$, and $P_3$ 
as exerted by the fluctuating gauge modes $a_{\mu }^{1}$,$a_{\mu }^{2}$, 
and $a_{\mu }^{3}$, respectively. One has
\begin{equation}
\label{partialpr}
P_{3}=2\frac{\pi ^{2}}{90}T^{4}\,,\ \text{ }P_{1}=P_{2}=-6\,T\int_{0}^{\infty }
\frac{k^{2}dk}{2\pi ^{2}}\ln \left( 1-e^{-\frac{\sqrt{%
m^{2}+k^{2}}}{T}}\right)\,.
\end{equation}
Notice that in deriving Eq.\,(\ref{partialpr}) the vacuum part 
in the one-loop expressions for the partial pressures  
can safely be neglected by virtue of momentum constraints in the effective theory 
\cite{Hofmann}, see also below. The total one-loop 
pressure $P_{\tiny\mbox{1-loop}}$ (exerted by fluctuating modes) is given by 
$P_{\tiny\mbox{1-loop}}=P_{1}+P_{2}+P_{3}$. 

The temperature evolution of 
the coupling $e$ is determined by the relation
\eqb
\label{alam}
a=2\pi e\lambda^{-3/2}
\eqe
and by the (inverted) 
solution of the (one-loop) evolution equation
\eqb
\label{evequ}
\pd_a\lambda=-\frac{24\lambda^4 a}{2\pi^6}\frac{D(2a)}{1+\frac{24\lambda^3 a^2}{2\pi^6}D(2a)}
\eqe
where 
\eqb
\label{Da}
D(a)\equiv \int_0^\infty dx\frac{x^2}{\sqrt{x^2+a^2}}\,\frac{1}{\exp(\sqrt{x^2+a^2})-1}\,,
\eqe
$\lambda\equiv\frac{2\pi T}{\Lambda}$, and $a\equiv \frac{m}{2T}$. Eq.\,(\ref{evequ}) 
guarantees the invariance of the Legendre transformations between thermodynamical 
quantities when going from the fundamental to the effective theory. 
 The evolution of $e$ with 
temperature exhibits a logarithmic pole, $e\propto -\log(\lambda-\lambda_c)$, 
where $\lambda_c=13.89$ denotes the critical value of the (dimensionless) temperature, and 
a plateau setting in for $\lambda$ slightly larger than $\lambda_c$. 
The value of $e$ at the plateau is $e=\sqrt{8}\pi\sim 8.89$.

In calculating radiative corrections to the free-quasiparticle (one-loop) 
pressure in a real-time formulation the loop momenta in the effective theory are subject to constraints. 
The latter emerge due to the existence of a maximal scale $|\phi|$ of 
resolution\footnote{Since the length scale $|\phi|^{-1}$ is deep inside the 
saturation regime for the spatial coarse-graining \cite{Hofmann,HerbstHofmann2004} 
physical quantities should not depend on a mild rescaling of $|\phi|$. We have checked the validity of this 
assertion when computing the polarization tensor of the massless mode \cite{SHG2006}.}. 
In \cite{Hofmann2006} these constraints were discussed, 
and we content ourselves with simply quoting them here. 
First, any propagating gauge mode with four-momentum $p$ cannot be further off its 
mass-shell then $|\phi|^2$. That is
\eqb
\label{cond1}
|p^2-m^2|\le |\phi|^2\,\ \ \ (\mbox{for a massive mode})\,,\ \ \ \ \ 
|p^2|\le |\phi|^2\,\ \ \ (\mbox{for a massless mode})\,.
\eqe
Second, the momentum transfer within a vertex needs to be 
constrained. For a three-vertex (ii) is already contained in (i) 
by momentum conservation in the vertex. For 
a four-vertex one needs to distinguish $s$, $t$, and $u$ 
channels in the scattering process. Suppose that the ingoing (outgoing) momenta 
are labeled by $p_1$ and $p_2$ ($p_3$ and $p_4=p_1+p_2-p_3$). 
Then the following three conditions emerge
\eab
\label{cond2}
|(p_1+p_2)^2|&\le&|\phi|^2\,,\ \ \ (s\ \mbox{channel})\ \ \ \ \ \ \ \ \
|(p_3-p_1)^2|\le|\phi|^2\,,\ \ \ (t\ \mbox{channel})\nonumber\\ 
|(p_2-p_3)^2|&\le&|\phi|^2\,,\ \ \ (u\ \mbox{channel})\,.
\eae
For a three-vertex conditions (\ref{cond2}) are already contained in (\ref{cond1}) 
by momentum conservation in the vertex. Notice that the three 
conditions in Eq.\,(\ref{cond2}) reduce to 
the first condition if one computes the one-loop 
tadpole contribution to the polarization tensor or the two-loop contribution 
to a thermodynamical quantity, say the pressure, arising from a four-vertex \cite{HerbstHofmannRohrer2004,SHG2006}. 
Namely, the $t$-channel condition is then trivially satisfied 
while the $u$-channel condition reduces to the $s$-channel condition by letting 
the loop momentum $k\to -k$ in $|(p-k)^2|\le |\phi|^2$, 
see \cite{Hofmann,SHG2006,HerbstHofmannRohrer2004}. Notice also 
that upon a euclidean rotation $p_0\to ip_0$ 
the first condition in(\ref{cond1}) goes over in
\eqb
\label{cond1E}
|p^2+m^2|\le |\phi|^2\,.
\eqe
For SU(2) the quasiparticle mass is given as $m=2e\,|\phi|$ with $e\ge\sqrt{8}\pi\sim 8.89$ 
\cite{Hofmann}. Thus condition (\ref{cond1E}) is never satisfied, and massive 
modes propagate on-shell only. Conditions (\ref{cond1}) and 
(\ref{cond2}) imply that the higher the loop order the more suppressed their 
contribution to a thermodynamical quantity. General arguments 
suggest that, apart from diagrams associated with one-particle irreducible 
resummations of propagators, only a finite number of diagrams 
contributes to the loop expansion \cite{Hofmann2006}. 
The main purpose of the present 
work is to demonstrate the validity of this on the three-loop level.  

\section{Irreducible three-loop diagrams\label{T3D}}

We are interested in an estimate for the modulus of each irreducible three-loop 
diagram contributing to the pressure. By three-loop irreducible we mean 
that the diagram does not include any line that is dressed by (multiple) 
insertions of one-loop polarizations. These one-particle 
reducible contributions to the propagator must be resummed 
to avoid the occurrence of pinch singularities \cite{Hofmann2006,SHG2006}. 
Apart from a mild modification of the tree-level propagator this modifies 
the dispersion law of the 
associated mode which in turn leads to a slight  
modifications of the constraints in (\ref{cond1}). The claim of 
\cite{Hofmann2006} is that the loop expansion 
terminates with respect to irreducible diagrams, that is, with respect to all 
those diagrams which do not yield a 
one-particle reducible diagram upon performing a cut (in all possible ways) 
on a single line. 

\begin{figure}
\begin{center}
\leavevmode
\leavevmode
\vspace{4.3cm}
\includegraphics{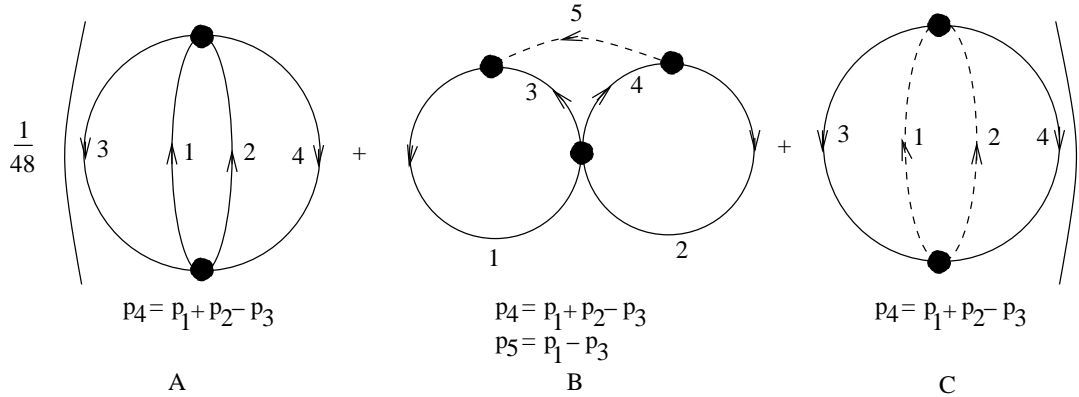}
\end{center}
\caption{\protect{\label{Fig-1}} Irreducible three-loop contributions to the pressure. Solid (dashed) 
lines are associated with the propagators of massive (massless) modes.}      
\end{figure}

The only three-loop irreducible diagrams are depicted in Fig\,\ref{Fig-1}. 
In the following we use the convention as in Fig\,\ref{Fig-1} for 
labelling the loop momenta. For diagrams A, B, and C 
the number $\tilde{K}$ of independent, potentially noncompact 
loop variables $(p_0,|\vec{p}|)_i,\ (i=1,2,3),$ is $\tilde{K}=6$, and the number $K$ of 
independent constraints is $K=7$. This implies that the support for 
the loop integrations is either compact or empty, see also the 
discussion in \cite{Hofmann2006}. As we shall see, the former 
possibility applies to diagrams A and B while diagram C vanishes.      

Let us first discuss diagrams A and B. Upon use of the Feynman rules, see 
\cite{SHG2006}, considering the symmetry factor $\frac{1}{48}$, 
by appealing to the triangle inequality, using the fact that 
the modes $a_{\mu}^{1,2}$ propagate (on-shell) thermally only, integrating over 
the time components of the independent loop momenta (momentum conservation), and 
after a rescaling of the radial components of loop momenta as 
\eqb
|\vec{p_i}|\rightarrow x_i\equiv \frac{|\vec{p_i}|}{|\phi|}\,, \ \ \ (i=1,2,3)
\eqe
one arrives at the following estimate for the 
moduli of the pressure corrections $\Delta P_{A}, \Delta P_{B}$ 
\eab
\label{DAB}
\left|\Delta P_{A(B)}\right|&\le&\frac{e^4\Lambda^4\lambda^{-2}}{3\times2^7\times(2\pi)^6}\sum^2_{l,m,n=1}\int dx_1\int dx_2\int
dx_3\int dz_{12}\int dz_{13}\int_{z_{23,l}}^{z_{23,u}} dz_{23}
\nonumber\\ 
&&\frac{1}{\sqrt{(1-z^2_{12})(1-z^2_{13})-(z_{23}-z_{12}z_{13})^2}}\frac{x_1^2x_2^2x_3^2}
{\sqrt{x_1^2+4e^2}\sqrt{x_2^2+4e^2}\sqrt{x_3^2+4e^2}}
\times\nonumber\\ 
&&\delta\left(4e^2+(-1)^{l+m}\sqrt{x_1^2+4e^2}\sqrt{x_2^2+4e^2}-
x_1x_2z_{12}-\right.\nonumber\\ 
&&\left.((-1)^{l+n}\sqrt{x_1^2+4e^2}\sqrt{x_3^2+4e^2}-
x_1x_3z_{13})-\right.\nonumber\\ 
&&\left.((-1)^{m+n}\sqrt{x_2^2+4e^2}\sqrt{x_3^2+4e^2}-
x_2x_3z_{23})\right)\times\nonumber\\ 
&&\left|{\cal
P}_{A(B)}(\vec{x},\vec{z},l,m,n)\right|\,n_B\left(2\pi\lambda^{-3/2}\sqrt{x_1^2+4e^2}\right)\times\nonumber\\ 
&&n_B\left(2\pi\lambda^{-3/2}\sqrt{x_2^2+4e^2}\right)
\,n_B\left(2\pi\lambda^{-3/2}\sqrt{x_3^2+4e^2}\right)\times\nonumber\\  
&&n_B\left(2\pi\lambda^{-3/2}\left|(-1)^{l}\sqrt{x_1^2+4e^2}+(-1)^{m}\sqrt{x_2^2+4e^2}+
(-1)^{n}\sqrt{x_3^2+4e^2}\right|\right)\,,\nonumber\\  
\eae
where $z_{12}\equiv\cos\angle(\vec{x}_1,\vec{x}_2)$, $z_{13}\equiv\cos\angle(\vec{x}_1,\vec{x}_3)$, 
and $z_{23}\equiv\cos\angle(\vec{x}_2,\vec{x}_3)$. The functions ${\cal P}_{A}$, ${\cal P}_{B}$ emerge from 
Lorentz and color contractions and are 
regular at $x_1=x_2=x_3=0$ (mass gap for $a_{\mu}^{1,2}$). We refrain from 
quoting them here explicitly for environmental reasons\footnote{Upon request the reader will be provided with the Mathematica 
notebooks containing the functions ${\cal P}_{A}$, ${\cal P}_{B}$ in explicit form.}. 
In addition, we define:
\eqb
\label{zlu}
z_{23,u}\equiv\cos\left|\arccos z_{12}-\arccos z_{13}\right|\,,\ \ \ 
z_{23,l}\equiv\cos\left|\arccos z_{12}+\arccos z_{13}\right|\,.
\eqe
The integrations in Eq.\,(\ref{DAB}) are subject to the 
following constraints (see (\ref{cond1}) and (\ref{cond2})):
\eab
\label{cond2b}
z_{12}&\le&\frac{1}{x_1x_2}\left(4e^2-\sqrt{x_1^2+4e^2}\sqrt{x_2^2+4e^2}+\frac{1}{2}\right)\equiv
g_{12}(x_1,x_2)\,,\nonumber\\ 
z_{13}&\ge&\frac{1}{x_1x_3}\left(-4e^2+\sqrt{x_1^2+4e^2}\sqrt{x_3^2+4e^2}-\frac{1}{2}\right)\equiv
g_{13}(x_1,x_3)\,,\nonumber\\ 
z_{23}&\ge&\frac{1}{x_2x_3}\left(-4e^2+\sqrt{x_2^2+4e^2}\sqrt{x_3^2+4e^2}-\frac{1}{2}\right)\equiv
g_{23}(x_2,x_3)\,.
\eae
Notice that for diagram B the constraint (\ref{cond1}) for 
the momentum $p_5=p_1-p_3$ of the massless mode is the same as the 
$t$-channel constraint for the four-vertex. Therefore no extra 
condition for the (off-shell) momentum $p_5$ is needed. 
According to the investigation in \cite{Hofmann2006} the conditions (\ref{cond2b}) 
together with Eq.\,(\ref{zlu}) imply that the support for the integration in $x_1,x_2,$ 
and $x_3$ is contained in the compact set $\{x_1,x_2,x_3<3\}$. 

Let us now turn to diagram C. We first consider the case that the momenta $p_1$ and 
$p_2$ of the massless modes, compare Fig.\,\ref{Fig-1}, 
are both off-shell within the constraints dictated by (\ref{cond1}). In analogy to diagrams 
A and B one then derives that
\eab
\label{DC}
\left|\Delta P_{C}\right|&\le&\frac{e^4\Lambda^4\lambda^{-2}}{3\times2^5\times(2\pi)^8}\sum^2_{l,m=1}
\int dy_1\int dx_1\int dx_2\int
dx_3\int dz_{12}\int dz_{13}\int_{z_{23,l}}^{z_{23,u}} dz_{23}
\nonumber\\ 
&&\frac{x_1^2x_2^2x_3^2}{\sqrt{(1-z^2_{12})(1-z^2_{13})-(z_{23}-z_{12}z_{13})^2}}
\,\left|{\cal P}_{C}(\vec{x},\vec{z},y_1,l,m)\right|\times\nonumber\\ 
&&n_B\left(2\pi\lambda^{-3/2}\sqrt{x_3^2+4e^2}\right)\,\frac{n_B
\left(2\pi\lambda^{-3/2}\left|(-1)^{l}\sqrt{x_3^2+4e^2}+(-1)^m f_2(\vec{x},\vec{z})\right|\right)}
{f_2(\vec{x},\vec{z})\sqrt{x_3^2+4e^2}}\,,\nonumber\\ 
\eae
where 
\eab
\label{defs}
f_2(\vec{x},\vec{z})&\equiv&
\sqrt{x_1^2+x_2^2+x_3^2+2x_1x_2z_{12}-2x_1x_3z_{13}-2x_2x_3z_{23}}\,,\nonumber\\ 
y_1&\equiv&\frac{p^0_1}{|\phi|}\,,
\eae
and $z_{23,l}$, ${z_{23,u}}$ are defined 
as in Eq.\,(\ref{zlu}). The function ${\cal P}_{C}$ emerges from 
Lorentz and color contractions and is 
regular at $x_1=x_2=x_3=0$ (mass gap for $a_{\mu}^{1,2}$). 
The integrations in Eq.\,(\ref{DC}) are subject to the 
following constraints
\eab
\label{ccC}
1&\ge&|y_1^2+y_2^2-x_1^2-x_2^2+2y_1y_2-2x_1x_2z_{12}|\,,\nonumber\\ 
1&\ge&|y_2^2-x_2^2+4e^2-(-1)^l 2y_2\sqrt{x_3^2+4e^2}+2x_2x_3z_{23}|\,,\nonumber\\ 
1&\ge&|y_1^2-x_1^2+4e^2-(-1)^l 2y_1\sqrt{x_3^2+4e^2}+2x_1x_3z_{13}|\,,\nonumber\\ 
1&\ge&|y_1^2-x_1^2|\,,\ \ \ \ \ \ \ \ 1\ge|y_2^2-x_2^2|\,,
\eae
where 
\eqb
\label{y2}
y_2=-y_1+2(-1)^l\,\sqrt{x_3^2+4e^2}+(-1)^m\,f_2(\vec{x},\vec{z})\,.
\eqe
As we shall see in Sec.\,\ref{res}, the constraints in 
(\ref{ccC}) imply that the support for the integration 
in Eq.\,(\ref{DC}) is empty. As a consequence, the cases 
that one or both of the massless modes in diagram C propagate 
on shell also have an empty support. This is 
because the conditions $|p_1^2|,|p_2^2|\le |\phi|^2$, which 
went into (\ref{ccC}), contain the cases $p_1^2=0$ 
and/or $p_2^2=0$. Thus diagram C is the 
first example of an vanishing 
irreducible diagram in the 
loop expansion. As was argued in \cite{Hofmann2006} 
one expects that the number of such cases will drastically 
increase with increasing loop numbers.

\section{Numerical evaluation and results\label{res}}

Here we present our results obtained by using the Monte-Carlo method 
of integration for the regular integrands\footnote{The $x_1$-integration is performed 
analytically in order to eliminate the $\delta$-function in the original integrand. 
There are eight zeros of the argument of the $\delta$-function in $x_1$ some of 
which turn out to be complex and thus can be discarded.} of 
Eqs.\,(\ref{DAB}) and (\ref{DC}). Let us first discuss diagrams A and B. 
It is known \cite{Hofmann2006} that the support in $x_2,x_3$ for the integral in Eqs.\,(\ref{DAB}) is 
contained in the compact set $\{x_2,x_3<3\}$ while the support for the integration in 
$z_{12},z_{13},z_{23}$ naturally is contained in the 
set $\{-1\le z_{12},z_{13}\le +1; z_{23,l}\le z_{23}\le z_{23,u}\}$, 
see Eq.\,(\ref{zlu}). Points are thus chosen 
randomly in the union of these two compact sets. 
Any point that satisfies the constraints (\ref{cond2b}) contributes 
to the integrals. We have worked with a sample size of $5\times10^5$ 
points, and we have observed a typical statistical uncertainty of 
about 1\% in our results. 
\begin{figure}
\begin{center}
\leavevmode
\leavevmode
\vspace{4.3cm}
\includegraphics{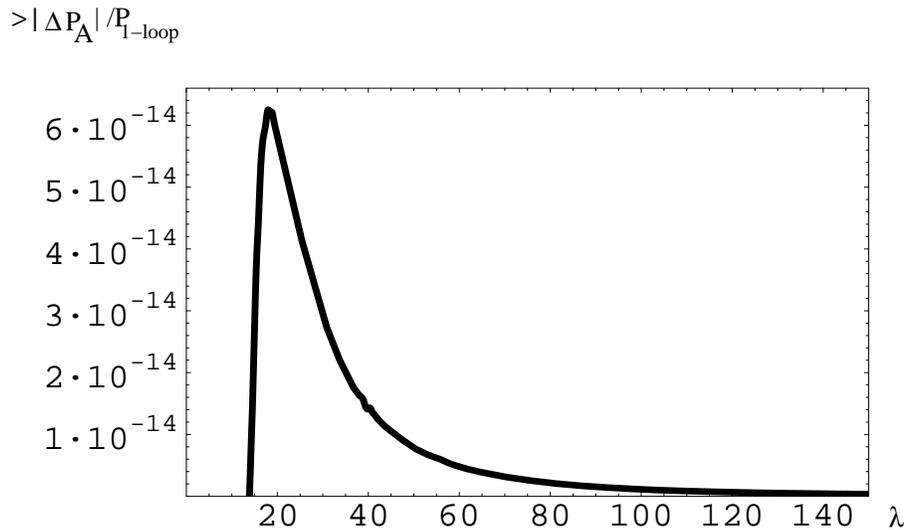}
\end{center}
\caption{\protect{\label{Fig-2}} An upper estimate for the modulus of 
the pressure contribution $|\Delta P_A|$ 
due to diagram A in Fig.\,\ref{Fig-1}. The plot shows 
this estimate normalized to the one-loop result $P_{\tiny\mbox{1-loop}}$, 
see Eq.\,(\ref{partialpr}). }      
\end{figure}
In Fig.\,\ref{Fig-2} our estimate for $\frac{|\Delta P_A|}{P_{\tiny\mbox{1-loop}}}$ 
is shown as a function of (dimensionless) temperature. Notice the sudden drop 
to zero near $\lambda_c=13.87$ which is due to the decoupling of the massive 
modes $a_{\mu}^{1,2}$. The functional shape is similar to 
that of the modulus of the leading two-loop correction: There is a maximum at 
$\lambda\sim 20$ and a very rapid decay to the right of this maximum. Notice, however, 
that the value of the maximum is suppressed by a factor of about $10^{-7}$ as compared to 
the smallest two-loop correction \cite{SHG2006}.  
\begin{figure}
\begin{center}
\leavevmode
\leavevmode
\vspace{5.7cm}
\includegraphics{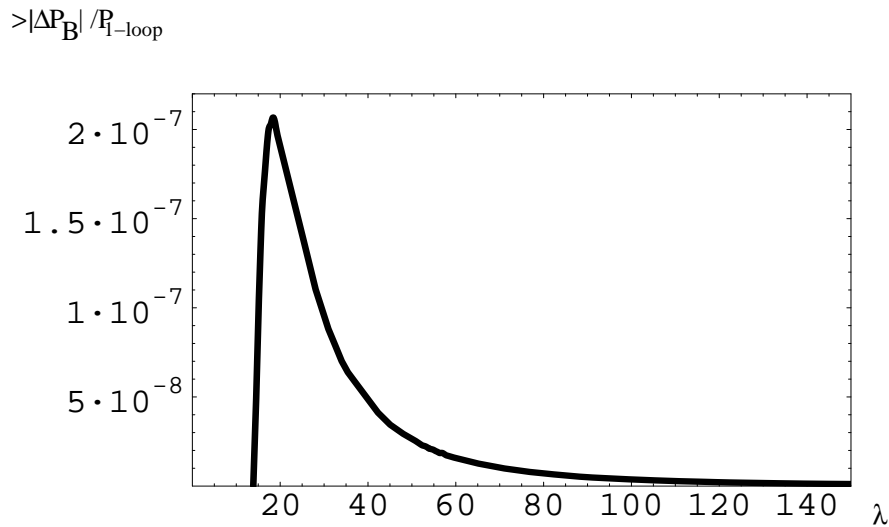}
\end{center}
\caption{\protect{\label{Fig-3}} An upper estimate for the modulus of 
the pressure contribution $|\Delta P_B|$ 
due to diagram A in Fig.\,\ref{Fig-1}. The plot shows 
this estimate normalized to the one-loop result $P_{\tiny\mbox{1-loop}}$, 
see Eq.\,(\ref{partialpr}). }      
\end{figure}
In Fig.\,\ref{Fig-3} we present our estimate for $\frac{|\Delta P_B|}{P_{\tiny\mbox{1-loop}}}$ 
as a function of (dimensionless) temperature. The maximum of this contribution is comparable to the 
smallest two-loop correction. 

For diagram C we have chosen a compact set $\{x_1,x_2,x_3\le R,-R\le y_1\le R,
-1\le z_{12},z_{13}\le +1; z_{23,l}\le z_{23}\le z_{23,u}\}$ in which the 
Monte-Carlo method samples points. We have varied $R$ in the range 
$0.1\le R\le 15$ and have used samples with up to $6\times10^8$ 
points. We have not found any point which satisfies all of 
the conditions (\ref{ccC}). This is physically suggestive since the diagram describes 
annihilation or creation of two massive on-shell modes into or out of two 
massless off-shell modes. Typically, the off-shellness of a 
massless modes is comparable to the mass of the massive mode. 
This very fact, however, is in stark contradiction
with condition (\ref{cond1}). At loop order three we thus have found a first 
example of an irreducible loop diagram that vanishes. 
We expect that this situation occurs very frequently at 
higher loop orders, see discussion in \cite{Hofmann2006}.       

\section{Conclusions\label{con}}

In this paper we have performed estimates on the moduli 
of the three irreducible three-loop diagrams which contribute to the pressure of 
a thermalized SU(2) Yang-Mills theory being in its deconfining phase. Our results are 
consistent with the general arguments in \cite{Hofmann2006} which imply a 
rapid convergence of the loop expansion. Namely, one of these three diagrams vanishes 
exactly because the constraints on the loop momenta, which emerge in the 
effective theory, imply that the support for the 
integration is empty. So the situation that an 
irreducible diagram is precisely zero, which in \cite{Hofmann2006} is argued to 
occur at a finite loop order, takes place at three-loops for the 
first time. Moreover, we observe that the 
modulus of the dominating three-loop diagram is comparable to that of the 
smallest two-loop diagram and by a
factor of $\sim 10^{-4}$ suppressed as compared to the 
dominating two-loop diagram \cite{SHG2006}. The modulus of the other nonvanishing, 
irreducible three-loop diagram is by a factor of $\sim 10^{-11}$ suppressed as compared to the 
dominating two-loop diagram.     

\section*{Acknowledgments}
We would like to thank Markus Schwarz for useful conversations and helpful comments on the 
manuscript. One of us (R.H.) acknowledges fruitful 
discussions with Francesco Giacosa.

\end{document}